\documentclass[floatfix,twocolumn,notitlepage,pra,  superscriptaddress]{revtex4-2}
\usepackage{academicons}
\usepackage{graphicx}
\usepackage{braket}
\usepackage{amsmath,amsfonts,amssymb,amstext,amscd,amsthm,bbold}
\usepackage{comment,float}

\usepackage{xcolor}
\usepackage{orcidlink}
\usepackage{hyperref}
\hypersetup{
    colorlinks=true,
    linkcolor=blue,
    filecolor=red,      
    urlcolor=teal,citecolor=purple
}

\definecolor{orcidlogocol}{HTML}{A6CE39}
\usepackage{silence}
\usepackage[normalem]{ulem}
\newcommand\numberthis{\addtocounter{equation}{1}\tag{\theequation}}

\renewcommand{\selectlanguage}[1]{}
\begin{document}
	\title{Open system approach to Neutrino oscillations in a quantum walk framework}
	
	
	\author{Himanshu Sahu$^{\orcidlink{0000-0002-9522-6592}}$}
	\email{himanshusah1@iisc.ac.in}
	\affiliation{Dept. of Instrumentation \& Applied Physics, Indian Institute of Sciences, C.V. Raman Avenue, Bengaluru 560012, India}
	\author{C. M. Chandrashekar$^{\orcidlink{0000-0003-4820-2317}}$}
	\email{chandracm@iisc.ac.in}
	\affiliation{Dept. of Instrumentation \& Applied Physics, Indian Institute of Sciences, C.V. Raman Avenue, Bengaluru 560012, India}
	\affiliation{The Institute of Mathematical Sciences, C. I. T. Campus, Taramani, Chennai 600113, India}
	\affiliation{Homi Bhabha National Institute, Training School Complex, Anushakti Nagar, Mumbai 400094, India}

	\begin{abstract}
		Quantum simulation provides a computationally-feasible approach to model and study many problems in chemistry, condensed-matter physics, or high-energy physics where quantum phenomenon define the systems behaviour. In high-energy physics, quite a few possible applications are investigated in the context of gauge theories and their application to dynamic problems, topological problems, high-baryon density configurations, or collective neutrino oscillations. In particular, schemes for simulating neutrino oscillations are proposed using a quantum walk framework. In this study, we approach the problem of simulating neutrino oscillation from the perspective of open quantum systems by treating the position space of quantum walk as environment.We have obtained the recurrence relation for Kraus operator which is used to represent the dynamics of the neutrino flavor change in the form of reduced coin states. We establish a connection between the dynamics of reduced coin state and neutrino phenomenology, enabling one to fix the simulation parameters for a given neutrino experiment and reduces the need for extended position space to simulate neutrino oscillations. We have also studied the behavior of linear entropy as a measure of entanglement between different flavors in the same framework.   \\
  \vspace{3mm}
  

	\end{abstract}

	\maketitle
	\section{Introduction}
	
	Quantum simulation\,\cite{RevModPhys.86.153} has become a rapidly growing field due to extraordinary development reporting the demonstration of controlling the quantum states in experimental setups\,\cite{browaeys_many-body_2020,hartmann_quantum_2016,houck_-chip_2012,greiner_quantum_2002,blatt_quantum_2012,monroe_programmable_2021,aspuru-guzik_photonic_2012,gross_quantum_2017,White:16}. Apart from the quantum advantages that comes from exploiting the laws of quantum mechanics\,\cite{cirac_goals_2012,daley_practical_2022}, it allows the study of the system from a quantum information theory perspective\,\cite{verstraete_density_2004,nalewajski_understanding_2019,wasielewski_exploiting_2020,lewis-swan_dynamics_2019,augusiak_many_2012,goold_role_2016,lloyd_quantum_2008}. There are two ways to approach problems via quantum simulation: digital or analog quantum computers. A digital quantum computer manipulates the quantum mechanical state using a series of discrete gate operations and can possibly perform error correction on imperfect operations. On the other hand, analog quantum computers are built to simulate the dynamics of a particular system or quantum algorithm. These have a significant advantage that they can be scaled to large system sizes. There are multiple experimental platforms where analog quantum simulations are realized, including neutral atoms, superconducting systems, trapped ions, and photons\cite{browaeys_many-body_2020,hartmann_quantum_2016,houck_-chip_2012,greiner_quantum_2002,blatt_quantum_2012,monroe_programmable_2021,aspuru-guzik_photonic_2012,gross_quantum_2017,White:16}.

	Quantum walk, the quantum mechanical analog of the classical random walks\,\cite{venegas-andraca_quantum_2012,ambainis,aharonov,ambainis2004quantum}, is a widely successful framework for modeling controlled dynamics in quantum systems\,\cite{oka_breakdown_2005,engel_evidence_2007,mohseni_environment-assisted_2008,PhysRevA.78.022314,chandrashekar_disordered-quantum-walk-induced_2011,kitagawa_exploring_2010}, and for building quantum algorithms\,\cite{ambainis2004quantum,shenvi_quantum_2003,Childs_2003,ambainis2004coins}. There are two broad categories of quantum walks - discrete-time quantum walk and continuous-time quantum walk. 
	
	In a discrete-time quantum walk, a quantum particle or a qudit is allowed to move in discrete steps on a lattice governed by unitary evolution operators. At each step, the particle state undergoes a unitary transformation which can be expressed as a product of two operators, namely the coin and the shift operators. The coin operator acts as a rotation in the qubit space and the shift operator translates the particle to another vertex on the lattice\,\cite{chandrashekar_discrete-time_2010,venegas-andraca_quantum_2012}. In case of continuous-time quantum walk, the quantum particle is described by a quantum state that evolves over time according a time-varying unitary operator. As opposed to discrete-time quantum walk, the continuous-time quantum walk evolution is continuous in time\,\cite{mulken_continuous-time_2011}. Experimental implementations of discrete-time quantum walks have been realized with lattice-based quantum systems where position space is mapped onto discrete lattice sites as well as circuit-based quantum processors\,\cite{flurin_observing_2017,ramasesh_direct_2017,qiang_efficient_2016,peruzzo_quantum_2010,tamura_quantum_2020,karski_quantum_2009,broome_discrete_2010,perets_realization_2008,zahringer_realization_2010,ryan_experimental_2005,schreiber_photons_2010}.

	In past developments, variants of discrete-time quantum walk have been used to simulate Dirac equation and its associated dynamics\,\cite{chandrashekar_two-component_2013,mallick_Dirac_2016,mallick_simulating_2019,PhysRevA.73.054302} which are also experimentally implemented in both analog and digital quantum simulators\, \cite{huerta_alderete_quantum_2020,gerritsma_quantum_2010}. These experimental implementation allows the laboratory realization and investigation of a variety of key fundamental phenomena associated with Dirac particle dynamics like Zitterbewegung, simultaneous position and spin oscillations\, \cite{gerritsma_quantum_2010,qu_observation_2013,lovett_observation_2023,liu_quantum_2014}. Schemes for quantum simulations of Dirac equation in curved space-time, discrete gauge theories, free quantum field theory, collective neutrino oscillations have been also reported in literature\, \cite{mallick_simulating_2019,brun_quantum_2020,arnault_quantum_2016,arrighi_quantum_2016,arnault_quantum_2016,arnault_discrete_time_2019,PhysRevD.104.063009}.

	\textit{Neutrino oscillations} is an interesting phenomenon in which new physics is to be expected from non-standard effects. In this work, we deal with flavor oscillations in neutrinos. Experiments and observations have established that the neutrino energy (mass) state does not coincide with the weak interaction (flavor) state. These two types of states are, in fact, related to one another by a unitary transformation. As a result, a neutrino produced in one flavor state can be detected in another flavor state at a later time. Neutrino oscillations have given rise to vibrant phenomena involving supernovae, reactor neutrinos, the early universe, or atmospheric neutrinos, and the solution to the solar neutrino problem\,\cite{lesgourgues_neutrino_2014,qian_physics_2019,hayes,huaiyu,mirizzi_supernova_2016,bahcall_solar_2004,garcia_phenomenology_2008}.
	
	A framework to simulate neutrino oscillation with discrete-time quantum walks has been proposed in previous literature\,\cite{mallick_neutrino_2017,molfetta_quantum_2016} and also studied from quantum information perspective\,\cite{banerjee_quantum-information_2015}. In this work, we approach the problem from the open quantum system perspective by considering the walker's evolution in the reduced coin space, thereby effectively treating the position space as environment. We consider the reduced dynamics of the coin state obtained by tracing over the position space degrees of freedom of a quantum walker. The Kraus operators we have obtained exhibit a temporal recurrence relation which allows one to calculate them at any given time systematically. This was not possible in earlier framework proposed for simulating neutrino oscillations. These Kraus operators are shown to describe the dynamics of the Dirac particle. We extended this formalism to describe the dynamics of more than one Dirac particle, thereby establishing a connection between the dynamics of reduced coin state and neutrino phenomenology. We studied the behavior of linear entropy as a measure of entanglement between different flavors in the same framework. The Kraus operators form presented in this work can be used as a guiding framework to model dynamics in other quantum systems where quantum walks and Dirac equations are used for its simulating and modelling the dynamics.

	This paper is organized as follows: In Sec.~\ref{sec:neutrino_oscillations}, we briefly discuss the basic theoretical background of neutrino oscillations. In Sec.~\ref{sec:DTQW}, we introduce discrete-time quantum walk formalism and describe its relation to the Dirac equation. We considered the reduced dynamics of the coin by tracing out the position space and derive the Kraus operators. We also find a recurrence relation for the Kraus operator in time. In Sec.~\ref{sec:passagefromtomorethanoneparticle}, we extended the formalism of quantum walks to describe the dynamics of more than one Dirac particle. In Sec.~\ref{sec:DTQW_with_Kraus}, we consider the quantum circuit construction of mixing matrix \eqref{eq:mixing_matrix} and propose a formalism for simulating neutrino oscillation with Kraus operators. In Sec.~\ref{sec:Results}, we present the numerical results and study the behavior of linear entropy as a measure of entanglement. We conclude in Sec.~\ref{sec:conclusion} with closing remarks. Throughout this paper, we use natural units, defined by $\hbar = c = 1$.

	\section{Neutrino Oscillations}\label{sec:neutrino_oscillations}
	
	In the theory of electroweak interaction developed by Glashow, Weinberg, and Salam, lepton flavor is conserved, and neutrinos are massless. Therefore, a neutrino of a given flavor created in charged current weak interactions will remain in the same flavor state. However, various experimental observations have shown that the flavor of neutrino changes as it propagates in space-time \cite{kajita_observation_2001,PhysRevLett.20.1205,gonzalez-garcia_phenomenology_2008}. This disconnection between the theoretical expectation and experimental observation is resolved by a scheme that relies on mixing the three neutrino mass and flavor states. According to this, each of the three neutrino flavor states is a mixture of the three mass eigenstate. As neutrinos move in space-time, each mass eigenstate acquires a different phase. Hence, a definite flavor state evolves into a mixture of three flavors, leading to flavor oscillation called \textit{neutrino oscillation} \cite{RevModPhys.59.671,bilenky_lepton_1978,bilenky_neutrino_2016}.

	Defining flavor eigenstates $|\nu_\alpha\rangle \ (\alpha = e,\mu, \tau)$ and mass eigenstates $|\nu_i\rangle \ (i=1,2,3)$, then the flavor and mass eigenstates are related by a unitary transformation written as
	\begin{equation}
		|\nu_\alpha \rangle = \sum_i U_{\alpha i} |\nu_i\rangle 
	\end{equation}
	
	where $U_{\alpha i}$ is mixing matrix known as the Pontecorvo-Maki-Nakagawa-Sakata (PMNS) matrix \cite{maki_remarks_1962} given by 
	\begin{equation}\label{eq:mixing_matrix}
		\begin{split}
			U&=\begin{bmatrix}U_{e1}&U_{e2}&U_{e3}\\U_{\mu 1}&U_{\mu 2}&U_{\mu 3}\\U_{\tau 1}&U_{\tau 2}&U_{\tau 3}\end{bmatrix}\\
			&=\begin{bmatrix}1&0&0\\0&c_{23}&s_{23}\\0&-s_{23}&c_{23}\end{bmatrix}\begin{bmatrix}c_{13}&0&s_{13}e^{-i\delta }\\0&1&0\\-s_{13}e^{i\delta }&0&c_{13}\end{bmatrix} \\
			& \qquad \qquad \times  \begin{bmatrix}c_{12}&s_{12}&0\\-s_{12}&c_{12}&0\\0&0&1\end{bmatrix}\begin{bmatrix}e^{i\alpha _{1}/2}&0&0\\0&e^{i\alpha _{2}/2}&0\\0&0&1\\\end{bmatrix} \\
			& \equiv U_3 U_2 U_1U_0
		\end{split}
	\end{equation}

	where $c_{ij} \equiv  \cos \phi_{ij}$, and $s_{ij} \equiv  \sin \phi_{ij}$ with $\phi_{ij}$ being the mixing angle. The phase factor $\alpha_1,\alpha_2$ are physically meaningful only if neutrinos are Majorana particles. The state $|\nu_i\rangle $ is the mass eigenstate of the free Dirac Hamiltonian (in natural units)
	\begin{equation}
		H_i  = \vec{\xi}\cdot \vec{p}_i + \beta m_i
	\end{equation}
	where $m_i$ is the mass, $\vec{p}_i$ is momentum operator. The propagation of eigenstate $|\nu_i\rangle$ can be described by plane wave solutions of the form 
	\begin{equation}
		|\nu_i(t)\rangle = e^{-i\left(E_i t - \vec{k}_i \cdot \vec{x} \right)}|\nu_i(0)\rangle 
	\end{equation}
	where $\vec{k}_i$ is the three dimensional momentum, with  $E_i  = \sqrt{|\vec{k}_i|^2  + m_i^2}$ being the positive energy of mass-eigenstate. 
	
	\vspace{3mm}
	\noindent
	Suppose that at time $t=0$ a flavor neutrino $|\nu_\alpha\rangle $ is produced, then at time $t$ the neutrino state given by 
	
	\begin{equation}\label{eq:state_evol}
		|\nu_\alpha (t)\rangle = \sum_i U_{\alpha i }|\nu_i(t)\rangle = \sum_i U_{\alpha i} e^{-iE_it} |\nu_i(0)\rangle.
	\end{equation}
	
	Therefore, the probability of transition $\nu_\alpha \rightarrow \nu_\beta $ after a time $t$ is given by 
	\begin{equation}\label{eq:transition_prob}
		P(\nu_\alpha\rightarrow \nu_\beta;t )= \left| \sum_j U^*_{\alpha j}U_{\beta j} e^{-iE_j t} \right|.
	\end{equation}
	
	To illustrate Eq\,\eqref{eq:transition_prob}, let us consider the simpler case with only two neutrino state. In this case, the mixing matrix $U$ can be written as 
	\begin{equation}
		U = \begin{bmatrix}
			\cos \phi & \sin \phi \\
			-\sin \phi & \cos \phi 
		\end{bmatrix}
	\end{equation}
	where $\phi $ is the mixing angle. In ultra-relativistic limit, $|\vec{k}_i| \gg m_i$ so that we can approximate energy as 
	\begin{equation}
		E_i =  \sqrt{|\vec{k}_i|^2 + m_i^2 } \simeq |\vec{k}_i|  + \frac{m_i^2 }{2|\vec{k}_i|} \approx E + \frac{m_i^2 }{2E}
	\end{equation} 
	where $E\approx |\vec{k}_i |$ for all $i$. The transition probability from flavor states $|\nu_\alpha\rangle $ to $|\nu_\beta\rangle $ in ultra-relativistic limit then, given by
	
	\begin{equation}\label{eq:transition_prob_two_flavor}
		P(\nu_\alpha \rightarrow \nu_\beta;t) = \sin^2 (2\phi )\sin^2 \left(\frac{\Delta m^2 L}{4E}\right)
	\end{equation}
	with $\Delta m^2= m_2^2 - m_1^2$ and $L \approx t$ is the distance traveled by the neutrinos from the production to the detection point.

	In atmospheric neutrino oscillations, the electron plays almost no role; hence Eq.\,\eqref{eq:transition_prob_two_flavor} is appropriate for flavor transition \cite{kajita_nobel_2016}. It is also appropriate for the solar case of $\nu_e \leftrightarrow \nu_X$ where $\nu_X$ is the superposition of $\nu_\mu$ and $\nu_\tau$. These approximations are possible since the mixing angle $\theta_{13}$ is very small and because two of the three mass eigenstates are very close in mass compared to the third \cite{fukuda_determination_2002}.

	\section{Discrete-time quantum walk}\label{sec:DTQW}

	The discrete-time quantum walk on a line is defined on a Hilbert space $\mathcal{H} = \mathcal{H}_c\otimes \mathcal{H}_p$
	where $\mathcal{H}_c$ is coin Hilbert space and $\mathcal{H}_p$ is the position Hilbert space. For a walk in one dimension, $\mathcal{H}_c$ is spanned by the basis set $|\uparrow\rangle $ and $|\downarrow\rangle$ representing the internal degree of the walker, and $\mathcal{H}_p$ is spanned by the basis state of the position $|x\rangle $ where $x\in \mathbb{Z}$ on which the walker evolves. At any time $t$, the state can be represented by 
	\begin{align}
		|\Psi(t)\rangle &= |\uparrow \rangle \otimes |\Psi^\uparrow(t)\rangle  + |\downarrow\rangle \otimes | \Psi^\downarrow(t)\rangle  \\
		&= \sum_x \begin{bmatrix}
			\psi^\uparrow_{x,t} \\
			\psi^\downarrow_{x,t}
		\end{bmatrix}.
	\end{align}
	Each step of the discrete-time quantum walk is defined by a unitary quantum coin operation $C$ on the internal degrees of freedom of the walker followed by a conditional position shift operation $S$ which acts on the configuration of the walker and position space. Therefore, the state at time $(t+\tau)$ where $\tau$ is the time required to implement one step of the walk will be 
	\begin{equation}
		|\Psi(t+\tau)\rangle = S(C\otimes I) |\Psi(t)\rangle= W|\Psi(t)\rangle.
	\end{equation}
	The general form of coin operator $C$, given by 
	\begin{equation}
		\begin{split}
			C &= C(\xi,\theta, \varphi ,\delta ) = e^{i\xi}e^{-i\theta \sigma_x}e^{-i\varphi \sigma_y}e^{-i\delta \sigma_z} \\
			&= e^{i\xi} \begin{pmatrix}
				e^{-i\delta}(c_\theta c_\varphi - is_\theta s_\varphi) & - e^{i\delta} (c_\theta s_\varphi + is_\theta c_\varphi) \\
				e^{-i\delta}(c_\theta s_\varphi - i s_\theta c_\varphi ) &e^{i\delta} (c_\theta c_\varphi + i s_\theta s_\varphi) 
			\end{pmatrix} \\
			&=  e^{i\xi} \begin{pmatrix}
				F_{\theta,\varphi , \delta} & G_{\theta, \varphi, \delta } \\
				-G^*_{\theta, \varphi, \delta } & F^*_{\theta,\varphi , \delta}
			\end{pmatrix}
		\end{split}
	\end{equation}
	where $c= \cos$, $s =\sin $, $\xi$ is global phase angle, $2\theta,2\varphi, 2\delta$ are the angles of rotations along $x,y$ and $z$ axes respectively with $\theta,\varphi,\delta \in [0,2\pi]$, and $\sigma_i$ is the $i$th component of the Pauli spin matrices $\{\sigma_x,\sigma_y,\sigma_z\}$, which are generators of SU(2) group. The position shift operator $S$ on the lattice with spacing $a$ is of the form
	\begin{equation}
		S = |\downarrow\rangle \langle \downarrow | \otimes T_+ + |\uparrow\rangle \langle \uparrow| \otimes T_- = \begin{bmatrix} 
			T_+ & 0\\
			0 & T_-
		\end{bmatrix}
	\end{equation}
	where 
	\begin{align*}
		T_\pm &= \sum_{x \in \mathbb{Z}} |x\pm a\rangle \langle x| 
	\end{align*}
	are translation operators. In momentum basis, these take a diagonal form 
	\begin{align*}
		T_\pm &= e^{\mp ipa} = \sum_k e^{\mp ika} |k\rangle \langle k | 
	\end{align*}
	with $|k\rangle $ being momentum eigenstate, given by
	\begin{equation}\label{eq:k_in_position_basis}
		|k\rangle = \frac{1}{\sqrt{2N+1}} \sum_{x=-N}^N e^{-ikx}|x\rangle
	\end{equation}
	
	where $k$ is eigenvalue which can take values $2\pi n/(2N+1)$, $n\in \mathbb{Z}$.  In ref.~\cite{chandrashekar_two-component_2013}, it is shown that for the coin operator 
	
	\begin{equation}\label{eq:C}
		\begin{split}
			B &\equiv C(0,\theta,0,3\pi/2)  \\
			&= \cos\theta |\uparrow\rangle \langle \uparrow |  + \sin\theta \left( |\uparrow\rangle \langle \downarrow | - |\downarrow\rangle \langle \uparrow |  \right)\\
			&\qquad \qquad \qquad \qquad + \cos \theta |\downarrow\rangle \langle \downarrow | 
		\end{split}
	\end{equation}
	
	the effective Hamiltonian defined as $W = e^{-iH \tau}$ takes the form of a one spatial dimensional Dirac Hamiltonian for small $k$ and $\theta$ with parameter correspondence given by 
	\begin{equation}\label{eq:par_corr}
		\frac{a}{\tau} = 1\ \ \ \ \&\ \ \ \ \frac{\theta}{\tau } = m.
	\end{equation}
	The eigenvector of $H$ corresponding to the positive eigenvalue 
	\begin{equation}
		E = \frac{1}{\tau} \arccos(\cos \theta \cos \tilde{k}) 
	\end{equation}
	where $\tilde{k} = ka$ and the corresponding eigenstates are given by
	\begin{align}
		|\nu\rangle = \begin{bmatrix}
			f(\theta,k) & g(\theta,k)
		\end{bmatrix}^T \otimes |k\rangle 
	\end{align}
	where 
	
	\begin{equation}
		\begin{split}
			f(\theta,k) &= \frac{\sin\theta e^{-ik}}{\sqrt{\sin^2\theta + \left[\cos \theta \sin k - \left(1- \cos^2 \theta \cos^2 k \right)^{1/2} \right]^2 }}  \\
			g(\theta ,k ) &= \frac{ i \left( \cos \theta \sin k - \left(1- \cos^2 \theta \cos^2 k \right)^{1/2}  \right)}{\sqrt{\sin^2\theta + \left[\cos \theta \sin k - \left(1- \cos^2 \theta \cos^2 k \right)^{1/2} \right]^2 }}.
		\end{split} 
	\end{equation}

	\subsection{Reduced dynamics of a coin}\label{subsec:Kraus_Op}
	In this section, we will consider the reduced dynamics of a coin in a DTQW, which previously has been studied in the context of non-markovian characteristics of the dynamics \cite{naikoo_non-markovian_2020}. Although, in contrast to previous work, we will derive a temporal recurrence relation for Kraus operator which allows one to find these at any time for a generic initial quantum state.

	Consider the initial state of the walker to be  
	\begin{equation}
		|\Psi(0)\rangle = |\chi\rangle \otimes |\psi_x\rangle.
	\end{equation}
	consisting of an internal spin state 
	\begin{equation}
		|\chi\rangle \in \mathcal{H}_c = \{a_\uparrow |\uparrow \rangle  + a_\downarrow |\downarrow\rangle : a_{\uparrow/\downarrow}\in \mathbb{C} \} 
	\end{equation}
	and a position state 
	\begin{equation}
		|\psi_x\rangle \in \mathcal{H}_p = \left\{\sum_{x\in \mathbb{Z}} c_x |x\rangle  : \sum_{x\in \mathbb{Z}}|c_x|^2<\infty  \right\}. 
	\end{equation}
	After $t$ steps, this becomes
	\begin{equation}
		|\Psi(t)\rangle = W^t |\Psi(0)\rangle = W^t\left( |\chi\rangle \otimes |\psi_x\rangle \right).
	\end{equation}
	The density matrix representation of the above state is given by 
	\begin{align}
		\rho(t) &= |\Psi(t)\rangle \langle \Psi(t) | \\
		&= W^t |\Psi(0)\rangle \langle \Psi(0)|(W^\dagger)^t.
	\end{align}
	Now, we can trace over the position space to get
	\begin{align*}
		\rho_c(t) &= \sum_{x=-t}^t \langle x | W^t |\Psi(0)\rangle \langle \Psi(0)|(W^\dagger)^t |x\rangle \\
		&= \sum_{x=-t}^t \langle x|W^t |\psi_x \rangle |\chi \rangle \langle \chi| \langle \psi_x |(W^\dagger)^t |x\rangle \\
		&= \sum_{x=-t}^t \tilde{K}_x(t) \rho_c(0) \tilde{K}^\dagger_x(t) \numberthis \label{eq:red_den}
	\end{align*}
	where 
	\begin{equation}
		\tilde{K}_x(t) \equiv  \langle x|W^t |\psi_x \rangle.
	\end{equation}
	For simplicity, let's take the initial state to be $|\psi_x\rangle = |0\rangle $ so that
	
	\begin{equation}\label{eq:def_Kraus}
		K_x(t) = \langle x |W^t |0\rangle.
	\end{equation}
	To make further progress, we break down the evolution operator $W$ in shift and coin operator as
	\begin{align*}
		W &= S(C\otimes I) \\
		&= \left[  |\uparrow\rangle \langle \uparrow| \otimes T_- + |\downarrow\rangle \langle \downarrow | \otimes T_+ \right] \left[C\otimes 1 \right] \\
		&=  |\uparrow\rangle \langle \uparrow| C \otimes T_- + |\downarrow\rangle \langle \downarrow |C \otimes T_+ \\
		&= C_\uparrow \otimes T_- + C_\downarrow \otimes T_+.
	\end{align*}
	where 
	\begin{equation}
		\begin{split}
			C_\uparrow &\equiv |\uparrow \rangle \langle \uparrow |C = e^{i\xi} \begin{pmatrix}
				F_{\theta,\varphi , \delta} & G_{\theta, \varphi, \delta } \\
				0 & 0
			\end{pmatrix} \\
			C_\downarrow & \equiv |\downarrow\rangle \langle \downarrow| C= e^{i\xi} \begin{pmatrix}
				0 & 0 \\
				-G^*_{\theta, \varphi, \delta } & F^*_{\theta,\varphi , \delta}
			\end{pmatrix}.
		\end{split}
	\end{equation}
	With this, consider Kraus operator at $t+1$
	\begin{align*}
		K_x(t+1) &= \langle x| W^{t+1}|0\rangle = \langle x | WW^t|0\rangle \\
		&= \sum_{x'} \langle x |W|x'\rangle \langle x'|W^t|0\rangle \\
		&= \sum_{x'} \langle x |W|x'\rangle K_{x'}(t) \numberthis \label{eq:Kraus_t1}.
	\end{align*}
	Now consider the first term in the expression 
	\begin{align*}
		\langle x |W|x'\rangle &= C_\uparrow \langle x|T_-|x'\rangle + C_\downarrow \langle x|T_+ |x'\rangle \\
		&= C_\uparrow \langle x|x'-a\rangle + C_\downarrow \langle x|x'+1\rangle \\
		&= C_\uparrow \delta_{x,x'-a} + C_\downarrow \delta_{x,x'+a}
	\end{align*}
	putting this into Eq.\,\eqref{eq:Kraus_t1}, we get
	\begin{align*}
		K_x(t+1)&= \sum_{x'} \left[ C_\uparrow \delta_{x,x'-a} + C_\downarrow \delta_{x,x'+a}\right] K_{x'}(t) \\
		&= C_\uparrow K_{x+a}(t) + C_\downarrow K_{x-a}(t).
	\end{align*}
	Therefore, we get a recurrence relation for the Kraus operator given by  
	
	\begin{equation}\label{eq:recc_rel}
		K_x(t+1) = C_\uparrow K_{x+a}(t) + C_\downarrow K_{x-a}(t).
	\end{equation}
	The initial Kraus operator at $t = 0$ given by, using the definition given in Eq.\,\eqref{eq:def_Kraus}
	\begin{align*}
		K_{x}(0) = \langle x|0\rangle  = \delta_{x,0}.
	\end{align*}
	If we start with initial position state to be $|x'\rangle $ then Kraus operators $K_{xx'}(t)$  are related to $K_x(t) $ by 
	\begin{equation}
		K_{xx'}(t) \equiv \langle x |W^t |x'\rangle =  K_{x-x'}(t)
	\end{equation}
	and therefore for a generic extended initial position state $|\psi_x\rangle =\sum_{x'} c_{x'} |x'\rangle  $, we have 
	\begin{equation}\label{eq:ext_Kraus}
		\tilde{K}_x(t) =\sum_{x'} c_{x'} K_{xx'}(t) = \sum_{x'} c_{x'} K_{x-x'}(t).
	\end{equation}

	Hence, for general position state $|\psi_x\rangle $, the Kraus operators are simply a linear combination of Kraus operators for initial state $|0\rangle $. 
	
	To illustrate Eq.\,\eqref{eq:recc_rel}, we derive the expression for the Kraus operator corresponding to coin operator in Eq.\,\eqref{eq:C} for initial position state $|0\rangle$. The operators $C_{\uparrow/\downarrow}$ for the coin operator \eqref{eq:C} have the form
	\begin{equation}
		\begin{split}
			B_\uparrow & \equiv |\uparrow \rangle \langle \uparrow | B =  \begin{pmatrix}
				\cos \theta & \sin \theta \\
				0 & 0
			\end{pmatrix} \\
			B_\downarrow & \equiv  |\downarrow \rangle \langle \downarrow | B = \begin{pmatrix}
				0 & 0 \\
				-\sin \theta & \cos \theta 
			\end{pmatrix}.
		\end{split}
	\end{equation}
	For a one-step walk, $t =1$
	\begin{equation}
		\begin{split}
			K_x(1) &= B_\uparrow K_{x+a}(0) + B_\downarrow K_{x-a}(0) \\
			&= B_\uparrow \delta_{x+a,0} + B_\downarrow \delta_{x-a,0}.
		\end{split}
	\end{equation}
	Hence, for a one-step walk, we have two Kraus operators given by 
	\begin{equation}
		\begin{split}
			K_{-a}(1) &= B_\uparrow =\begin{pmatrix}
				\cos \theta & \sin \theta \\
				0 & 0
			\end{pmatrix} \\ 
			K_{a}(1) &= B_\downarrow = \begin{pmatrix}
				0 & 0 \\
				-\sin \theta & \cos \theta 
			\end{pmatrix}.
		\end{split}
	\end{equation}
	For a two-step walk, $t=2$
	\begin{equation}
		\begin{split}
			K_x(2) &= B_\uparrow K_{x+a}(1) + B_\downarrow K_{x-a}(1) \\
			&= B_\uparrow \left(B_\uparrow K_{x+2a}(0) + B_\downarrow K_{x}(0) \right)\\
			& \qquad \qquad \qquad +  B_\downarrow\left(B_\uparrow K_{x}(0) + B_\downarrow K_{x-2a}(0)\right) \\
			&= B_\uparrow \left(B_\uparrow \delta_{x+2a,0} + B_\downarrow \delta_{x,0} \right) \\ 
			& \qquad \qquad \qquad + B_\downarrow\left(B_\uparrow \delta_{x,0} +B_\downarrow \delta_{x-2a,0}\right) \\
			&= B_\uparrow B_\uparrow \delta_{x+2a,0} + (B_\uparrow B_\downarrow + B_\downarrow B_\uparrow)\delta_{x,0} \\
			& \qquad \qquad \qquad + B_\downarrow B_\downarrow \delta_{x-2a,0}
		\end{split}
	\end{equation}
	Therefore, for a two-step walk, we have three Kraus operators, given by 
	\begin{equation}
		\begin{split}
			K_{-2a}(2) &= B_\uparrow B_\uparrow = \begin{pmatrix}
				\cos^2(\theta) & \sin(\theta) \cos(\theta) \\
				0 & 0
			\end{pmatrix} \\
			K_{0}(2) &=  B_\uparrow B_\downarrow + B_\downarrow B_\uparrow = \begin{pmatrix}
				-\sin^2(\theta)  & \sin(\theta) \cos(\theta) \\
				-\sin(\theta)\cos(\theta) & \sin^2(\theta)
			\end{pmatrix} \\
			K_{2a}(2) &= B_\downarrow B_\downarrow = \begin{pmatrix}
				0 & 0 \\
				-\sin(\theta)\cos(\theta) & \cos^2(\theta)
			\end{pmatrix} .
		\end{split}
	\end{equation}
	
	In the similar manner, we can find the Kraus operator for a $t$-steps walk using the recurrence relation \eqref{eq:recc_rel}.
	
	\subsection{Passage to more than one particle}\label{sec:passagefromtomorethanoneparticle}
	
	We have seen that for the coin operator in Eq.\,\eqref{eq:C}, the discrete-time quantum walk mimics Dirac equation, therefore describes the dynamics of Dirac particle, with parameter correspondence as in Eq.\,\eqref{eq:par_corr}. We can extend this coin operator to reproduce a set of Dirac equations, hence describing the dynamics of more than one Dirac particle. To this end, consider a discrete-time quantum walk with $2n$-dimensional coin Hilbert space $\mathcal{H}_c$ spanned by basis $\bigoplus_{f=1,n}\{|f,\uparrow \rangle ,|f,\downarrow\rangle \} $ and coin operator 
	
	\begin{equation}\label{eq:C_nparticle}
		B_n = \bigoplus_{f=1,n} B_f(\theta_f)
	\end{equation}
	where 
	\begin{equation}\label{eq:Cf}
		\begin{split}
			B_f &= \cos\theta_f |f,\uparrow\rangle \langle f,\uparrow |  + \sin\theta_f \left( |f,\uparrow\rangle \langle f,\downarrow | - |f,\downarrow\rangle \langle f,\uparrow |  \right)\\
			&\qquad \qquad \qquad \qquad + \cos \theta_f |f,\downarrow\rangle \langle f,\downarrow |.
		\end{split}
	\end{equation}
	In this basis, the evolution operator takes the block diagonal form given by 
	
	\begin{equation}
		W = S (B_n\otimes  I) = \bigoplus_{f=1,n} W_f =  \bigoplus_{f=1,n} S_f(B_f\otimes I)
	\end{equation}
	with 
	\begin{align}\label{eq:S_f}
		S_f &= T_+ \otimes |f,\uparrow\rangle \langle f,\uparrow | + T_- \otimes |f,\downarrow\rangle \langle f,\downarrow |.
	\end{align}
	Analogous to coin operator Eq.\,\eqref{eq:C}, coin operator in Eq.\,\eqref{eq:C_nparticle} reproduces a set of $n$ Dirac equations with parameter correspondence 
	\begin{equation}
		\frac{a}{\tau} = 1\ \ \ \ \&\ \ \ \ \frac{\theta_f}{\tau } = m_f,\ \ \ \ f= 1,\ldots ,n
	\end{equation}
	
	with $f$ being the particle number with internal degree of freedom $\{|f,\uparrow \rangle ,|f,\downarrow\rangle \}$ and mass $m_f$. The Kraus operator formalism that we developed in Sec.~\ref{subsec:Kraus_Op} can easily be extended for coin operator in Eq.\,\eqref{eq:C_nparticle} as follows

	\begin{align}\label{eq:Kraus_op_nparticle}
		\tilde{\mathcal{K}}^{(n)}_x(t) &= \bigoplus_{f=1,n} \langle x|W_f |\psi_x\rangle =\bigoplus_{f=1,n}  \tilde{K}_x(\theta_f,t).
	\end{align}
	
	where extended Kraus operator $\tilde{K}_x(\theta_f,t)$ follow from Eq.\,\eqref{eq:ext_Kraus}
	
	\begin{equation}
		\tilde{K}_x(\theta_f,t) = \sum_{x'}c_{x'}K_{x-x'}(\theta_f,t) 
	\end{equation}
	with $K_{x-x'}(\theta_f,t) = \langle x |W_f^t|x'\rangle.$ With the formalism for simulating dynamics of $n$ dirac particle at hand, in the next section, we will see how to simulate neutrino oscillations.

	\section{Simulating neutrino oscillations using reduced quantum walk dynamics on coin space}\label{sec:DTQW_with_Kraus}
	
	In this section, we will consider the simulation of neutrino oscillation. We begin with considering the quantum circuit construction associated with mixing matrix \eqref{eq:mixing_matrix} and then move on to simulation two-flavor neutrino oscillations and later generalize to the three flavor cases. 
	
	\subsection{Quantum circuit construction of PMNS Matrix}
	
	Firstly, we map the neutrino flavor states $|\nu_\alpha\rangle $ into a three-qubit system \cite{banerjee_quantum-information_2015} so that the correspondence between the two systems looks like : 
	\begin{equation}
		\begin{split}
			|\nu_e\rangle \rightarrow |100\rangle \qquad |\nu_\mu\rangle \rightarrow |010\rangle \quad \quad |\nu_\tau\rangle \rightarrow |001\rangle .
		\end{split}
	\end{equation}
	
	With this, we can also write the mass-eigenstates using the mixing matrix as 
	
	\begin{equation}
		|\nu_i\rangle = U^*_{ei} |100\rangle + U^*_{\mu i} |010\rangle + U^*_{\tau i} |001\rangle .
	\end{equation}

Consider now the representation of the mixing matrix on this basis. In the three-qubit system, the mixing matrix would be a $8\times 8 $ matrix. To see the explicit form of these, consider the term $U_1$ in mixing matrix \eqref{eq:mixing_matrix}
\begin{equation}
	U_1 = \begin{bmatrix}c_{12}&s_{12}&0\\-s_{12}&c_{12}&0\\0&0&1\end{bmatrix} =
	\begin{array}{c|ccc}
		& |100\rangle & |010\rangle & |001\rangle \\
		\hline
		|100\rangle & c_{12} & s_{12} & 0 \\
		|010\rangle & -s_{12} & c_{12} & 0 \\
		|001\rangle & 0 & 0 & 1
	\end{array}
\end{equation}

which has a three-qubit representation 

\begin{equation}
	U_1 \rightarrow \begin{bmatrix}
		1 & 0 & 0 & 0 & 0 & 0 & 0 & 0 \\
		0 & 1 & 0 & 0 & 0 & 0 & 0 & 0 \\
		0 & 0 & c_{12} & 0 & -s_{12} & 0 & 0 & 0 \\
		0 & 0 & 0 & 1 & 0 & 0 & 0 & 0 \\
		0 & 0 & s_{12} & 0 & c_{12} & 0 & 0 & 0 \\
		0 & 0 & 0 & 0 & 0 & 1 & 0 & 0 \\
		0 & 0 & 0 & 0 & 0 & 0 & 1 & 0 \\
		0 & 0 & 0 & 0 & 0 & 0 & 0 & 1 \\
	\end{bmatrix}.
\end{equation}

so that there's a mixing between the terms $|100\rangle$ and $|010\rangle$ while the other remains the same. In the similar manner, we can write a three-qubit representation for other terms $U_i (i= 0,1,2,3)$ as follows :

\begin{equation}
	U_0 \rightarrow \begin{bmatrix}
	1 & 0 & 0 & 0 & 0 & 0 & 0 & 0 \\
	0 & 1 & 0 & 0 & 0 & 0 & 0 & 0 \\
	0 & 0 & e^{i\alpha_2/2} & 0 & 0 & 0 & 0 & 0 \\
	0 & 0 & 0 & 1 & 0 & 0 & 0 & 0 \\
	0 & 0 & 0 & 0 & e^{i\alpha_1/2} & 0 & 0 & 0 \\
	0 & 0 & 0 & 0 & 0 & 1 & 0 & 0 \\
	0 & 0 & 0 & 0 & 0 & 0 & 1 & 0 \\
	0 & 0 & 0 & 0 & 0 & 0 & 0 & 1 \\
\end{bmatrix}
\end{equation}

\begin{equation}
	U_2 \rightarrow \begin{bmatrix}
		1 & 0 & 0 & 0 & 0 & 0 & 0 & 0 \\
		0 & c_{13} & 0 & 0 & -s_{13}e^{i\delta} & 0 & 0 & 0 \\
		0 & 0 & 1 & 0 & 0 & 0 & 0 & 0 \\
		0 & 0 & 0 & 1 & 0 & 0 & 0 & 0 \\
		0 & s_{13}e^{-i\delta} & 0 & 0 & c_{13} & 0 & 0 & 0 \\
		0 & 0 & 0 & 0 & 0 & 1 & 0 & 0 \\
		0 & 0 & 0 & 0 & 0 & 0 & 1 & 0 \\
		0 & 0 & 0 & 0 & 0 & 0 & 0 & 1 \\
	\end{bmatrix}
\end{equation}

\begin{equation}
	U_3 \rightarrow \begin{bmatrix}
		1 & 0 & 0 & 0 & 0 & 0 & 0 & 0 \\
		0 & c_{23} & -s_{23} & 0 & 0 & 0 & 0 & 0 \\
		0 & s_{23} & c_{23} & 0 & 0 & 0 & 0 & 0 \\
		0 & 0 & 0 & 1 & 0 & 0 & 0 & 0 \\
		0 & 0 & 0 & 0 & 1 & 0 & 0 & 0 \\
		0 & 0 & 0 & 0 & 0 & 1 & 0 & 0 \\
		0 & 0 & 0 & 0 & 0 & 0 & 1 & 0 \\
		0 & 0 & 0 & 0 & 0 & 0 & 0 & 1 \\
	\end{bmatrix}.
\end{equation}

The unitary operations $U_i$ can be thought of as controlled unitary operators. To illustrate this, consider on our earlier example of $U_3$, which is a controlled operation on the third qubit (counting from left in $|ijk\rangle$). To see this, we can rewrite the unitary $U_3$ as :

\[
\begin{array}{c|cccccccc}
	& 000 & \textcolor{red}{001} & 010 & \textcolor{red}{011} & 100 & \textcolor{red}{101} & 110 & \textcolor{red}{111} \\
	\hline 
	000 & 1 & 0 & 0 & 0 & 0 & 0 & 0 & 0 \\
	\textcolor{red}{001} & 0 & 1 & 0 & 0 & 0 & 0 & 0 & 0 \\
	010 & 0 & 0 & c_{12} & 0 & -s_{12} & 0 & 0 & 0 \\
	\textcolor{red}{011} & 0 & 0 & 0 & 1 & 0 & 0 & 0 & 0 \\
	100 & 0 & 0 & s_{12} & 0 & c_{12} & 0 & 0 & 0 \\
	\textcolor{red}{101} & 0 & 0 & 0 & 0 & 0 & 1 & 0 & 0 \\
	110 & 0 & 0 & 0 & 0 & 0 & 0 & 1 & 0 \\
	\textcolor{red}{111} & 0 & 0 & 0 & 0 & 0 & 0 & 0 & 1 \\
\end{array}
\]

Or more explicitly, if we define a unitary matrix 

\begin{equation}
	\mathcal{U}_3 = \begin{bmatrix}
		1 & 0 & 0 & 0 \\
		0 & c_{12} & -s_{12} & 0 \\
		0 & s_{12} & c_{12} & 0 \\
		0 & 0 & 0 & 1
	\end{bmatrix}
\end{equation}

Then controlled operation look like : 
\begin{equation}
	\begin{split}
		|ij1\rangle &\rightarrow |ij1\rangle \\ 
		|ij0 \rangle & \rightarrow \mathcal{U}_3|ij\rangle \otimes |0\rangle    
	\end{split}
\end{equation}

The similar construction can be done for $U_0, U_2,U_3$.

	
	\subsection{Two-flavor neutrino oscillations}

	In case of two-flavor neutrino oscillations, we need to mimic the dynamics of two Dirac particles and therefore the coin hilbert space is four dimensional space spanned by $\bigoplus_{f=1,2} \{|f,\uparrow \rangle ,|f,\downarrow\rangle \}$. The evolution operator $W$ has a block diagonal form given by
	\begin{equation}\label{eq:Wblock}
		W = \bigoplus_{f=1,2} W_f = S(B_2\otimes I) = \bigoplus_{f=1,2} S_f( B_f \otimes I)
	\end{equation}
	
	where the quantum coin operator and shift operator are as in Eq.\,\eqref{eq:Cf} and Eq.\,\eqref{eq:S_f}. The mass eigenstates given by 
	
	\begin{equation}\label{eq:mass_eigenstate}
		\begin{split}
			|\nu_1\rangle &= \begin{bmatrix}
				f(\theta_1,k) & g(\theta_1,k) & 0 & 0
			\end{bmatrix}^T \otimes |k\rangle \equiv |\nu_1\rangle_c \otimes |k\rangle \\
			|\nu_2\rangle &= \begin{bmatrix}
				0 & 0 & f(\theta_2,k) & g(\theta_2,k) 
			\end{bmatrix}^T \otimes |k\rangle \equiv  |\nu_2\rangle_c \otimes |k\rangle
		\end{split}.
	\end{equation}

	The initial state $|\Psi(0)\rangle $ of the neutrino corresponding to $\alpha$ flavor using the mixing matrix acting on each sector
	\begin{equation}
		|\Psi(0)\rangle = |\nu_\alpha \rangle = \sum_{i=1,2} U_{\alpha i}|\nu_i\rangle.
	\end{equation}
	
	The associated reduced coin density matrix given by 
	\begin{equation}
		\rho_c(0) = \sum_{i,j} U_{\alpha i} U^*_{\alpha j} |\nu_i\rangle_c \langle \nu_j|_c.
	\end{equation}
	The Kraus operator for two particle given by  
	\begin{align*}
		\tilde{\mathcal{K}}_x(t) &= \bigoplus_{f=1,2} \langle x|W_f |\psi_x\rangle =\bigoplus_{f=1,2}  \tilde{K}_x(\theta_f,t)
	\end{align*}
	where state $|\psi_x\rangle $ is momentum eigenstate $k$ in position space representation \eqref{eq:k_in_position_basis}.
	At any time $t$, the reduced density matrix is written as 
	\begin{equation}\label{eq:density_mat2}
		\rho_c(t) = \sum_x \tilde{\mathcal{K}}^{(2)}_x(t) \rho_c(0) (\tilde{\mathcal{K}}^{(2)}_x)^\dagger(t).
	\end{equation}
	The probability of the $\nu_\alpha \rightarrow \nu_\beta$ transition after a time $t$ is then given by expectation value of the projection operator $|\nu_\beta \rangle_c \langle \nu_\beta |_c$ i.e.
	\begin{equation}\label{eq:DTQW_Prob}
		P(\nu_\alpha \rightarrow \nu_\beta ;t) = \text{Tr}\left[|\nu_\beta \rangle_c \langle \nu_\beta |_c \rho_c(t) \right]
	\end{equation}
	where 
	\begin{equation}
		|\nu_\beta\rangle_c = \sum_{i=1,2} U_{\beta i} |\nu_i\rangle_c
	\end{equation}
	so that 
	\begin{equation}
		|\nu_\beta\rangle_c \langle \nu_\beta|_c = \sum_{i,j=1,2}U_{\beta i }U^*_{\beta j} |\nu_i\rangle_c\langle \nu_j|_c.
	\end{equation}

	\subsection{Three-flavor neutrino oscillations}
	We can trivially extend the two-flavor neutrino oscillations to three-flavor neutrino oscillation by considering the coin operator, Eq.\,$\eqref{eq:C_nparticle}$ with $n = 3$. The mass eigenstates associated with three-particles are given by

	\begin{equation}\label{eq:mass_eigenstate3}
		\begin{split}
			|\nu_1\rangle &= \begin{bmatrix}
				f(\theta_1,k) & g(\theta_1,k) & 0 & 0 & 0 & 0
			\end{bmatrix}^T \otimes |k\rangle \equiv |\nu_1\rangle_c \otimes |k\rangle \\
			|\nu_2\rangle &= \begin{bmatrix}
				0 & 0 & f(\theta_2,k) & g(\theta_2,k) & 0 & 0 
			\end{bmatrix}^T \otimes |k\rangle \equiv  |\nu_2\rangle_c \otimes |k\rangle \\ 
			|\nu_2\rangle &= \begin{bmatrix}
				0 & 0 & 0 & 0 & f(\theta_3,k) & g(\theta_3,k)  
			\end{bmatrix}^T \otimes |k\rangle \equiv  |\nu_3\rangle_c \otimes |k\rangle \\ 
		\end{split}.
	\end{equation}
	
	The initial state $|\Psi(0)\rangle$ of the neutrino corresponding to $\alpha$ flavor using the mixing matrix acting on each sector 
	\begin{equation}
		|\Psi(0)\rangle = |\nu_\alpha \rangle = \sum_{i=1,2,3} U_{\alpha i} |\nu_i\rangle.
	\end{equation}
	The associated reduced coin density matrix given by 
	
	\begin{equation}
		\rho_c(0) = \sum_{i,j=1,2,3} U_{\alpha i} U^*_{\alpha j} |\nu_i\rangle_c \langle \nu_j|_c.
	\end{equation}
	The Kraus operator for three particle given by  
	\begin{align*}
		\tilde{\mathcal{K}}_x(t) &= \bigoplus_{f=1,2,3} \langle x|W_f |\psi_x\rangle =\bigoplus_{f=1,2,3}  \tilde{K}_x(\theta_f,t)
	\end{align*}
	where state $|\psi_x\rangle $ is momentum eigenstate $k$ in position space representation \eqref{eq:k_in_position_basis}.
	At any time $t$, the reduced density matrix is written as 
	\begin{equation}\label{eq:density_mat3}
		\rho_c(t) = \sum_x \tilde{\mathcal{K}}^{(3)}_x(t) \rho_c(0) (\tilde{\mathcal{K}}^{(3)}_x)^\dagger(t).
	\end{equation}
	The probability of the $\nu_\alpha \rightarrow \nu_\beta$ transition after a time $t$ is then given by expectation value of the projection operator $|\nu_\beta \rangle_c \langle \nu_\beta |_c$ i.e.
	\begin{equation}\label{eq:DTQW_Prob_Den}
		P(\nu_\alpha \rightarrow \nu_\beta ;t) = \text{Tr}\left[|\nu_\beta \rangle_c \langle \nu_\beta |_c \rho_c(t) \right]
	\end{equation}
	where 
	\begin{equation}
		|\nu_\beta\rangle_c = \sum_{i=1,2,3} U_{\beta i} |\nu_i\rangle_c
	\end{equation}
	so that 
	\begin{equation}
		|\nu_\beta\rangle_c \langle \nu_\beta|_c = \sum_{i,j=1,2,3}U_{\beta i }U^*_{\beta j} |\nu_i\rangle_c\langle \nu_j|_c.
	\end{equation}
	
	\section{Results}\label{sec:Results}
	
	In Sec.\,\ref{sec:DTQW_with_Kraus}, we have seen that evolution operator in Eq.\,\eqref{eq:Wblock} describes the set of Dirac equations that describes neutrino flavor oscillations and, we can make use this to establish a map with neutrino phenomenology, therefore allowing one to fix the QW parameters for a given neutrino experiment.
	
	\begin{figure}
		\centering
		\includegraphics[scale = 0.5]{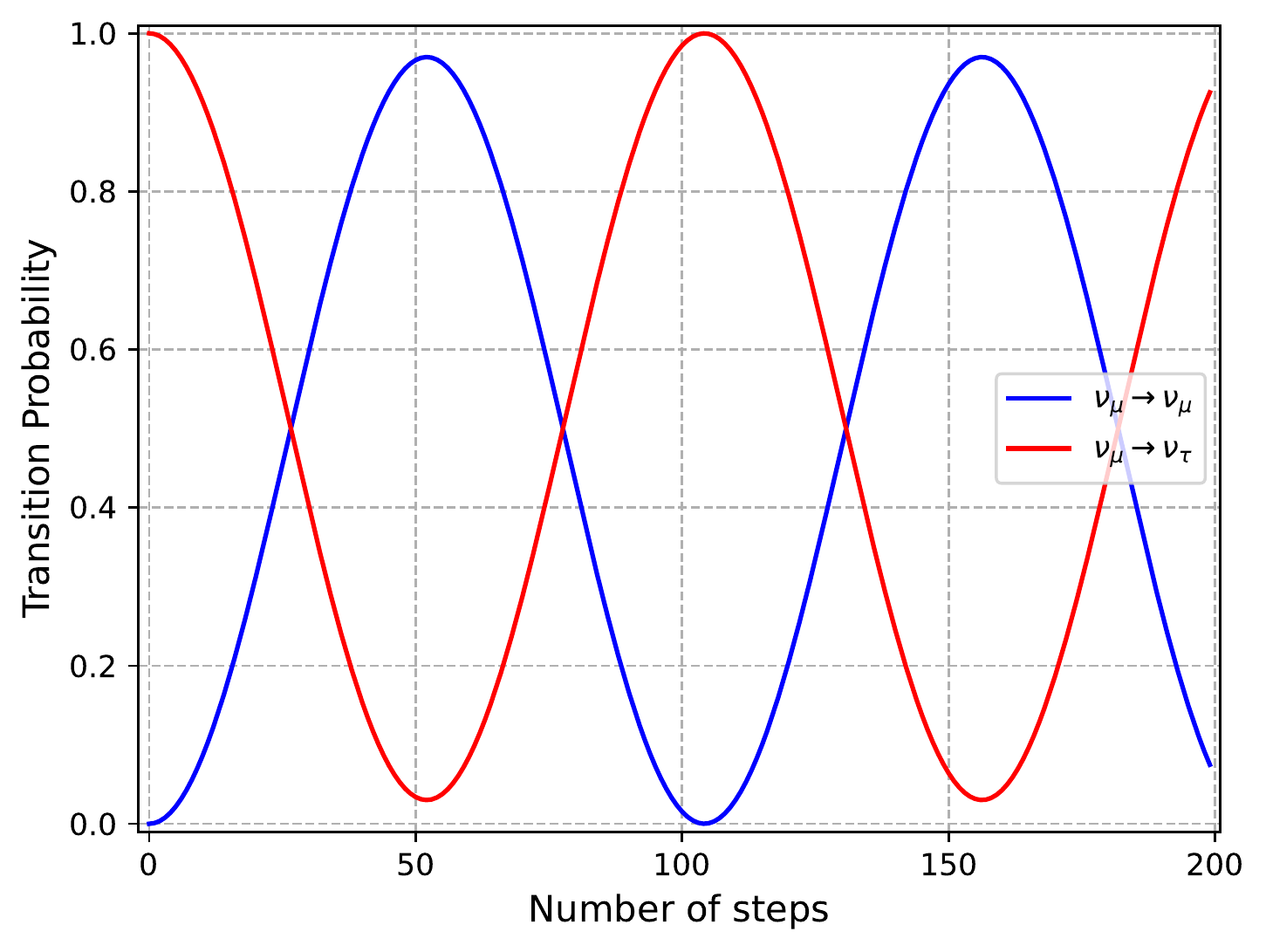}
		\caption{Transition probabilities of two flavor neutrino oscillation obtained from numerical simulation using the Kraus operator associated with the DTQW with initial state $|\nu_\mu\rangle$. The coin angles are  $\theta_1 =0.001$ rad., $\theta_2 = 0.0986$ rad., and the mixing angle $\phi= 0.698$ rad. with $\tilde{k} = 0.05$.}
		\label{fig:2Flavor}
	\end{figure}

\begin{figure}
		\centering
		\includegraphics[scale = 0.5]{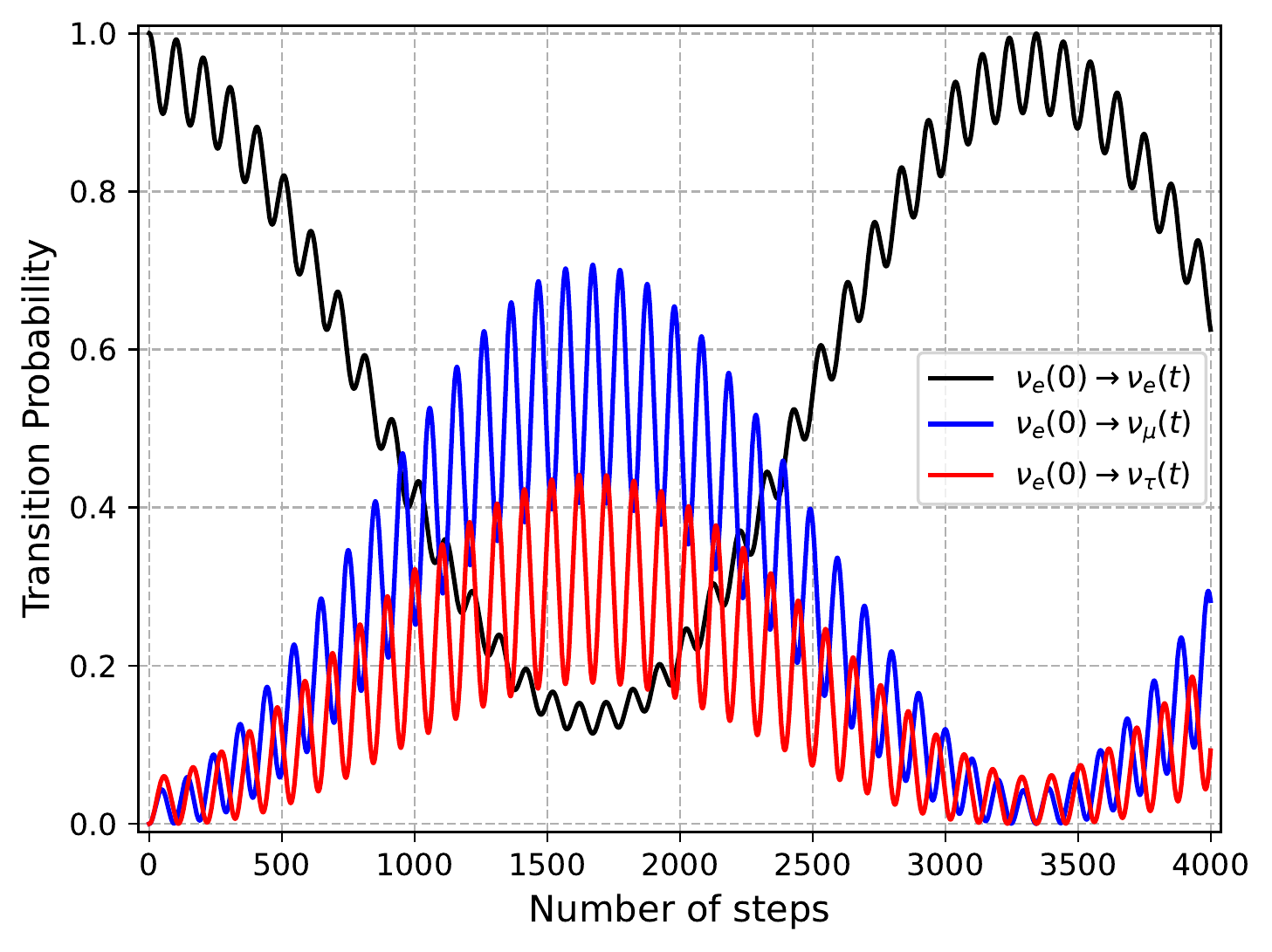}
		\includegraphics[scale = 0.5]{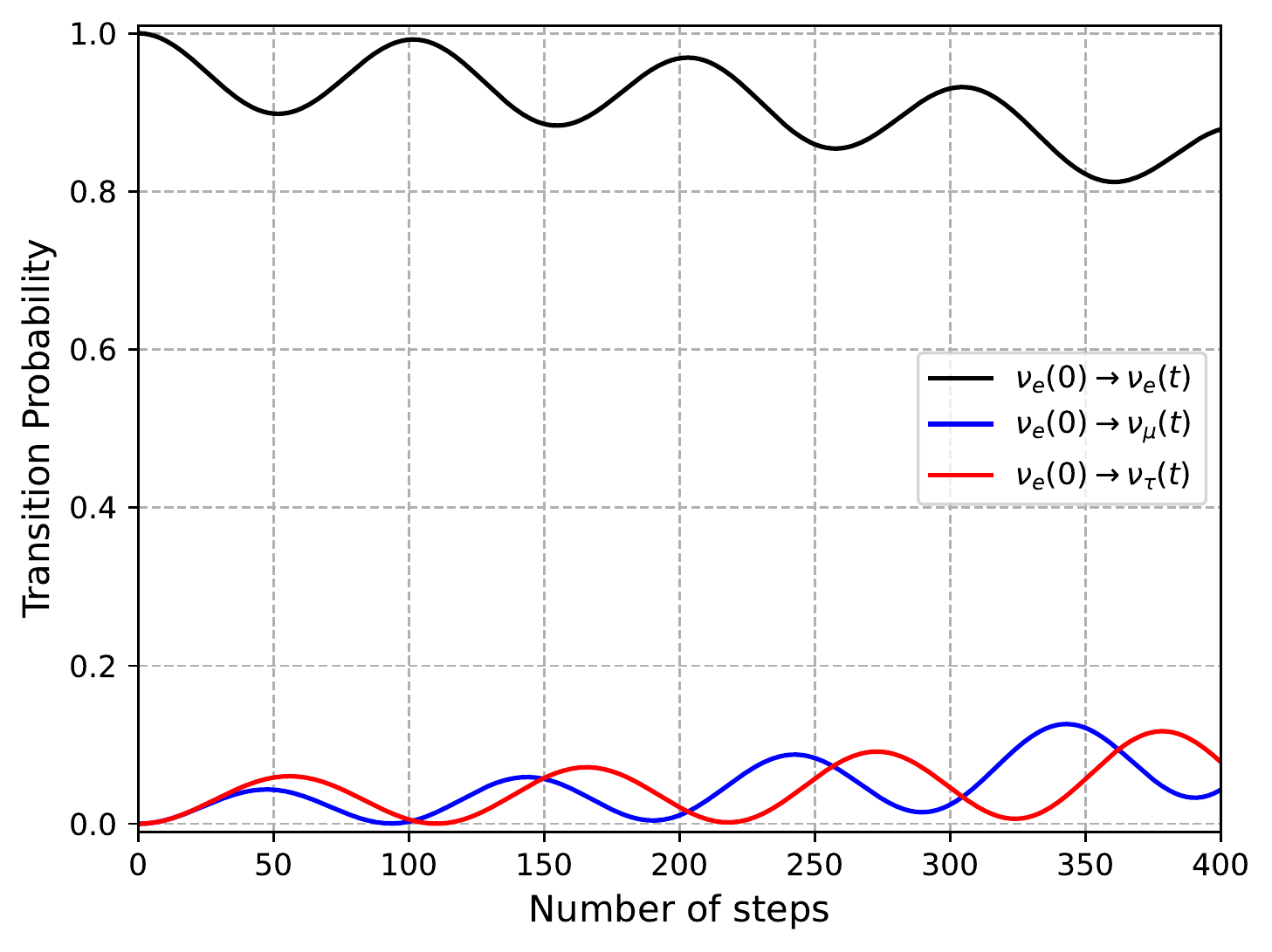}
		\caption{Transition probabilities of three flavor neutrino oscillation obtained from numerical simulation using the Kraus operator associated with the DTQW [a] for long time steps and [b] short time steps with initial state $|\nu_e\rangle.$. The coin angles are  $\theta_1 =0.001$ rad., $\theta_2 = 0.01963$ rad. $\theta_3 = 0.12797$ rad., and the mixing angle $\phi_{13} = 0.16087$ rad., $\phi_{23} =0.69835 $, $\phi_{12} =0.59437 $, and $\delta = 0$ with $\tilde{k} = 0.1$.}
		\label{fig:3Flavor}
	\end{figure}

	More explicitly, using the correspondence made between quantum walk and Dirac equation in Eq.\,\eqref{eq:par_corr}, we can write the oscillation frequency in Eq.\,\eqref{eq:transition_prob_two_flavor} in terms of quantum parameters as 
	\begin{equation}
		\begin{split}
			\frac{\Delta m^2_{ij} L}{4E} &= \frac{\Delta \theta^2_{ij}}{4\tilde{k}} \frac{t}{\tau}
		\end{split}
	\end{equation}
	where $\Delta \theta^2_{ij} = \theta_i^2 - \theta_j^2$ and $\Delta m^2_{ij} = m_i^2 - m_j^2$. Fig.~\ref{fig:2Flavor} and \ref{fig:3Flavor} shows the transition probability as a function of the number of steps of DTQW for two-flavor and three-flavor neutrino oscillations, respectively obtained from evolution describe by the Kraus operators. We can observe the oscillatory behavior of flavors. One can observe that the plot reproduces the one corresponding to actual calculations of neutrino oscillations (see for example \cite{wiki:Neutrino_oscillation}) with taken mass values 
	
	\begin{equation}
		\begin{split}
			\Delta m^2_{21} &= 7.50 \times 10^{-5} \ \text{eV}^2 \\
			\Delta m^2_{31} &= 2.457 \times 10^{-3} \ \text{eV}^2 \\
			\Delta m^2_{32} &= 2.382 \times 10^{-3} \ \text{eV}^2 \\
		\end{split}.
	\end{equation}

	\noindent
	{\it Linear entropy}:~Among various measures of degree of entanglement \cite{RevModPhys.80.517}, the linear entropy is shown to be useful in studying entanglement in neutrino oscillations \cite{blasone_entanglement_2009}. The linear entropy associated with a bipartite system is defined as 
	\begin{equation}
		S_j = \frac{d}{d-1}\left(1 - \text{Tr}(\rho_j^2)\right),\ \ \ \ j= 1,2.
	\end{equation}
	
	where $\rho_j$ represent the reduced density matrix of system $j$ and $d$ is dimension of density matrix $\rho_j$.

	In case of two flavor neutrino oscillations, we can write the state shown in Eq.\,\eqref{eq:state_evol} as the Bell-like superposition 
	
	\begin{equation}\label{eq:bell_state}
		|\nu_\alpha (t) \rangle = \tilde{U}_{\alpha \mu}(t) |1\rangle_{\nu_\mu}|0\rangle_{\nu_\tau} + \tilde{U}_{\alpha \tau} |0\rangle_{\nu_\mu}|1\rangle_{\nu_\tau} 
	\end{equation}
	where
	\begin{equation}\label{eq:bell_coeff}
		\tilde{U}_{\alpha \beta}(t) = \sum_{j } U_{\alpha j} U^*_{\beta j} e^{-iE_jt}.
	\end{equation}
	
	The linear entropies can found by straightforward calculation using associated density matrix of state $|\nu_\alpha (t)\rangle$, and are given by 
	\begin{equation}\label{eq:entropy}
		S_{\mu} = S_{\tau} = 4|\tilde{U}_{\alpha e}(t)|^2 |\tilde{U}_{\alpha \mu}(t)|^2.
	\end{equation}
	
\begin{figure}
		\centering
		\includegraphics[scale = 0.5]{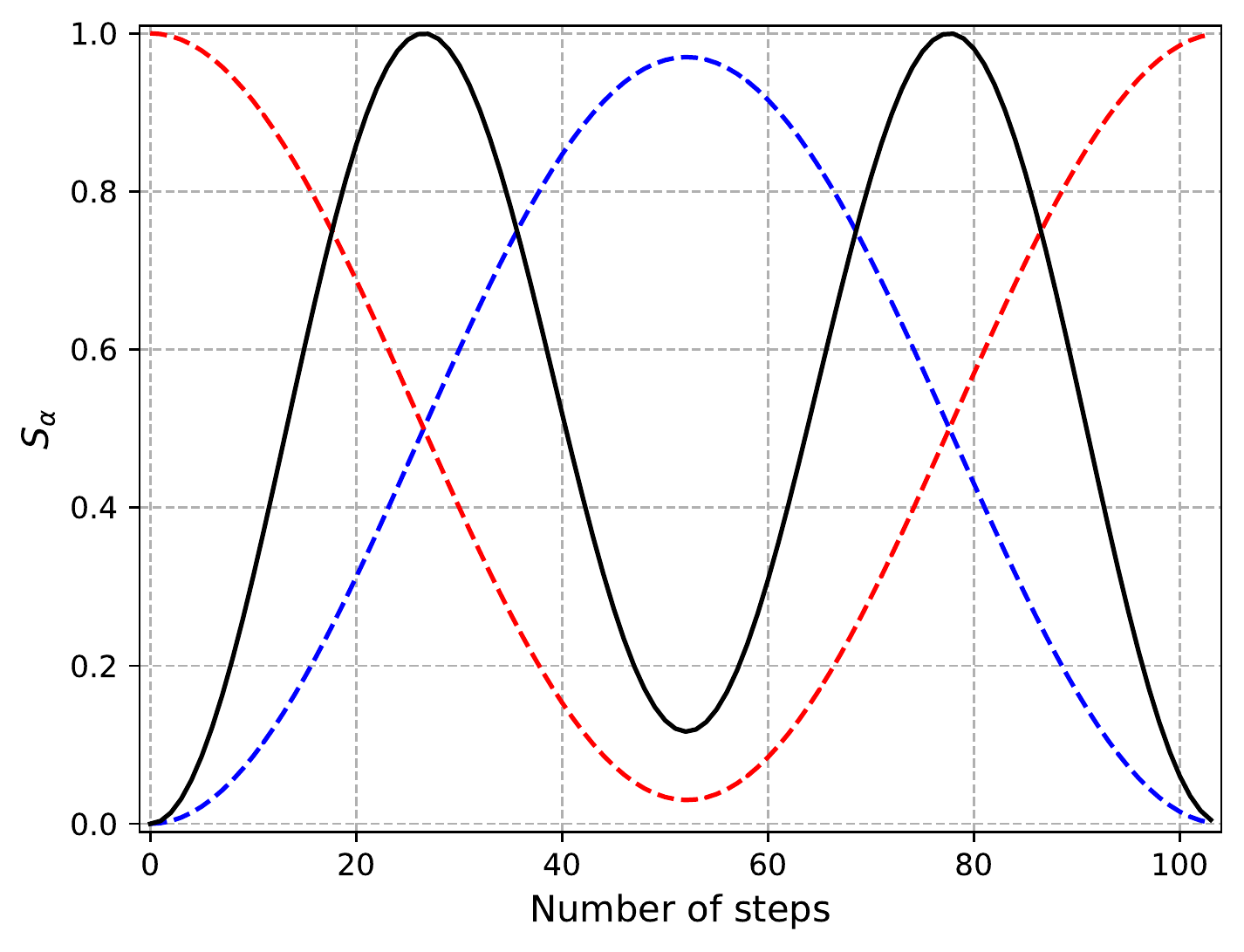}
		\caption{Linear entropy $S\mu$ (full line) as a function of number of steps (shown for single cycle) along with transition probabilities (dashed lines) $P(\nu_\mu \rightarrow \nu_\mu)$ (red) and $P(\nu_\mu\rightarrow \nu_\tau)$ (blue).   }
		\label{fig:Entropy}
	\end{figure}

	Therefore, the linear entropies can be written in term of transition probabilities which continues to be true for three flavor case \cite{blasone_entanglement_2009}. Fig.~\ref{fig:Entropy} shows the linear entropy $S_\mu$ obtained from transition probabilities given by Eq.\,\eqref{eq:DTQW_Prob} using Eq.\,\eqref{eq:entropy}. We find that degree of entanglement is largest when both transition probabilities are equals to $0.5$ which corresponds to maximally entangled Bell pair state and minimum when one of the transition probability is zero which corresponds to unentangled state.

	\vspace{3mm}
	
	For three-flavor case, we can write three-flavor state as 
	\begin{align*}
		|\nu_\alpha (t)\rangle &= \tilde{U}_{\alpha e}|1\rangle_{\nu_e}|0\rangle_{\nu_\mu}|0\rangle_{\nu_\tau} + \tilde{U}_{\alpha \mu}|0\rangle_{\nu_e}|1\rangle_{\nu_\mu}|0\rangle_{\nu_\tau} \\ 
		& \ \ \ + \tilde{U}_{\alpha \tau}|0\rangle_{\nu_e}|0\rangle_{\nu_\mu}|1\rangle_{\nu_\tau}. \numberthis
	\end{align*}

\begin{figure}
		\centering
		\includegraphics[scale = 0.35]{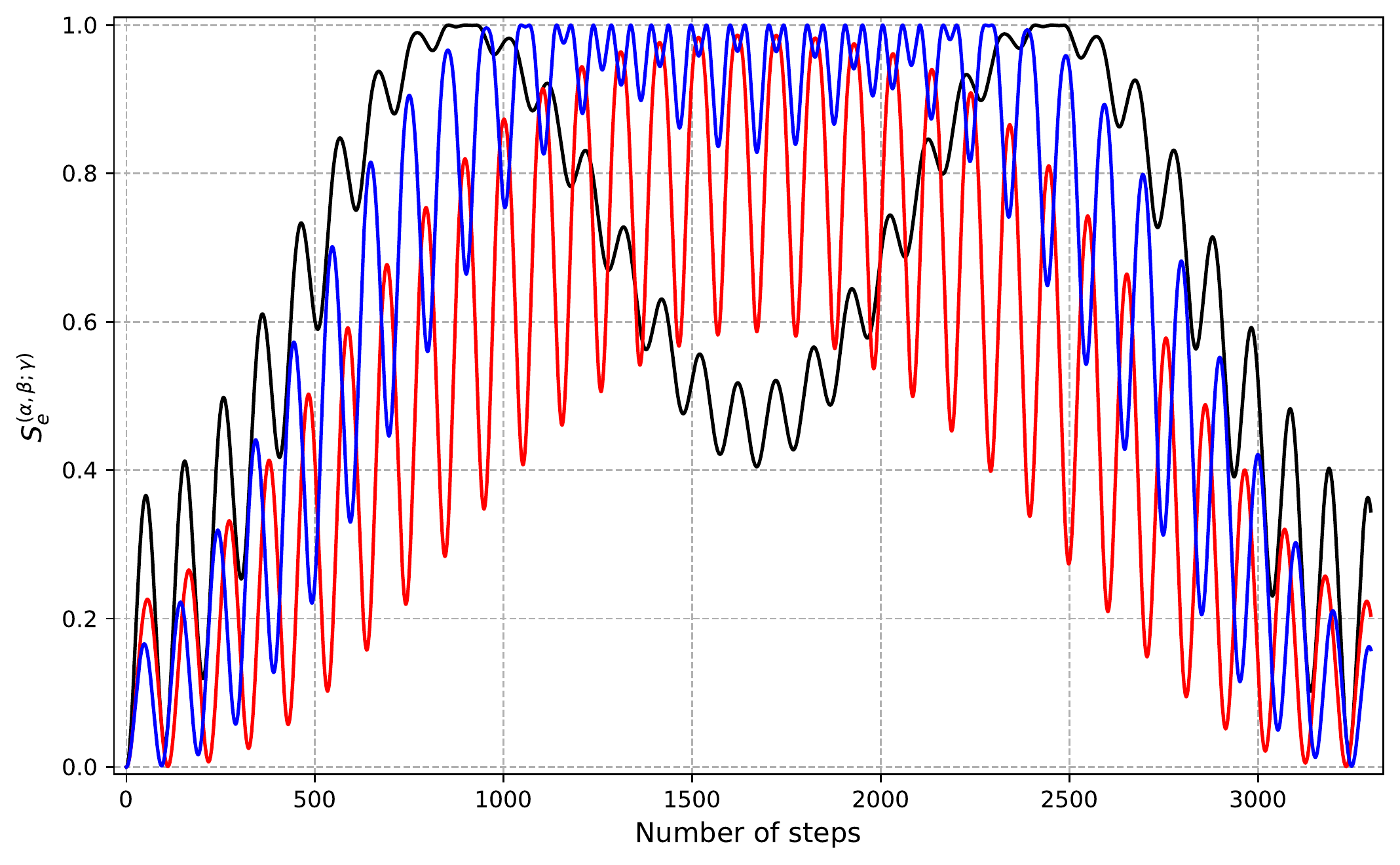}
		\caption{Partial linear entropies $S_e^{(\alpha,\beta;\gamma)}$ as a function of number of steps (shown for a single cycle) with black corresponds to $S_e^{(\mu,\tau;e)}$, blue corresponds to $S_e^{(\tau,e;\mu)}$, and red corresponds to $S_e^{(e,\mu;\tau)}$. }
		\label{fig:LEntropy3}
	\end{figure}

	In case of multipartite system, we can define partial linear entropies corresponding to various bipartition of total system \cite{blasone_entanglement_2009}. We adopt the notation $S_\zeta^{(\alpha,\beta;\gamma)}$ for linear entropy of reduced density matrix $\rho_\zeta^{(\alpha,\beta)} = \text{Tr}_\gamma(\rho_\zeta)$ where $\zeta$ corresponds to initial flavor state. Similar to two flavor case, the partial linear entropies can be written in terms of transition probabilities and given by 
	
	\begin{equation}
		S_\alpha^{(e,\mu;\tau)} = 4|\tilde{U}_{\alpha \tau}(t)|^2 \left( 1 -  |\tilde{U}_{\alpha \tau}(t)|^2 \right).
	\end{equation}
	The remaining two partial entropies can be found by simply permuting $e,\mu,\tau$. Fig.~\ref{fig:LEntropy3} shows the partial linear entropies for initial state $|\nu_e(0)\rangle$. The maximum of these partial shows the point at which maximum entanglement exist between those two flavors. To understand the behavior of overall degree of entanglement, we can define average linear entropy as mean value of partial entropies, given by 
	\begin{equation}
		\langle S_\alpha\rangle = \frac{8}{3} \left(|\tilde{U}_{\alpha e} |^2|\tilde{U}_{\alpha \mu} |^2  + |\tilde{U}_{\alpha e}(t)|^2|\tilde{U}_{\alpha \tau} |^2 +|\tilde{U}_{\alpha \mu } |^2|\tilde{U}_{\alpha e\tau} |^2 \right).
	\end{equation}

	\begin{figure}[t]
		\centering
		\includegraphics[scale = 0.35]{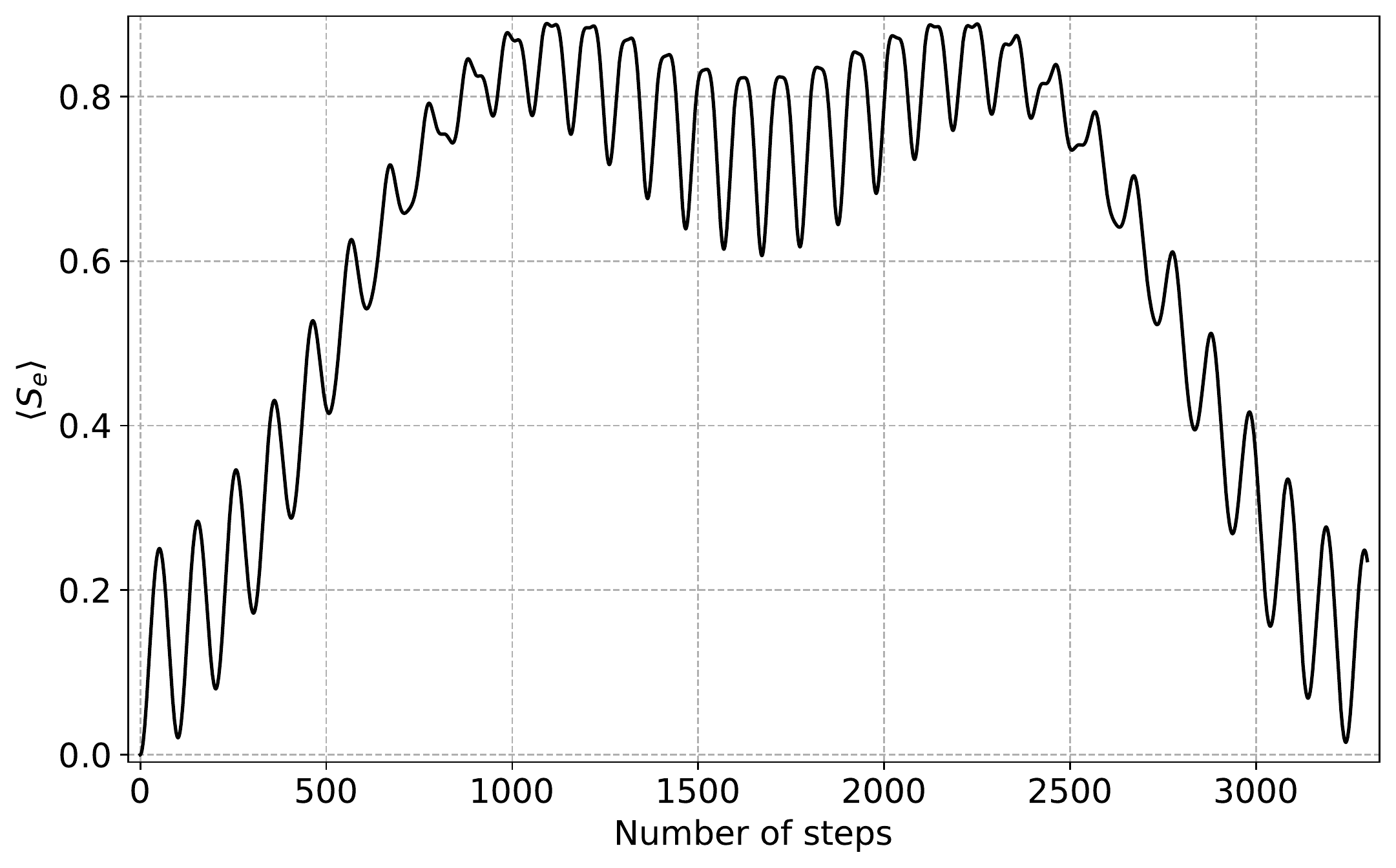}
		\caption{Average linear entropy $\langle S_e\rangle $ as a function of number of steps (shown for a single cycle).}
		\label{fig:AVLEntropy3}
	\end{figure}
	
	Fig.~\ref{fig:AVLEntropy3} shows behavior of average linear entropy $\langle S_e\rangle $, we observe similar behavior as in case of two-flavor neutrino oscillations. This shows strong correlation between the components $\nu_\mu$ and $\nu_\tau.$

	\section{Conclusion}\label{sec:conclusion}
	
	In this work, we proposed a novel scheme to simulate neutrino oscillations using DTQW formalism. We considered the evolution of reduced dynamics of coin density matrix using set of Kraus operators obtained from tracing out the position space. This in turn means effectively treating dynamics in position space as environmental effect. We obtained the transition probabilities of neutrino flavor states in the same framework. To study the degree of entanglement between different flavors, we considered linear entropy which found to be maximum for bell-pair state.

	We conclude with a few interesting future directions for this work. Firstly, given the recent development in simulating open system dynamics in quantum devices \cite{Hu2020,Hu2022generalquantum}, the demonstration of a quantum algorithm of open system approach to neutrino oscillation on a near-term quantum device would give laboratory verification of the phenomenon. Furthermore, in previous studies,  possible decoherence effects induced by new physics (e.g., quantum gravity, string theory) in neutrino oscillations have been studied by considering the open system framework \cite{Fabio_Benatti_2000}, and bounds on dissipative parameters are obtained for various neutrino experiments \cite{PhysRevLett.85.1166,PhysRevD.100.055023}. Our scheme is open to incorporating these dissipation effects, which may provide an exciting direction for investigating non-standard effects.

	\begin{acknowledgments}
		We would like to thank Mr. Prateek Chawla for his comments on the manuscript.
	\end{acknowledgments}
 \bibliographystyle{apsrev4-1}
\bibliography{ref}
	
\end{document}